\documentclass[sigconf]{acmart} 
\AtBeginDocument{%
  \providecommand\BibTeX{{%
    \normalfont B\kern-0.5em{\scshape i\kern-0.25em b}\kern-0.8em\TeX}}}
    
\renewcommand\footnotetextcopyrightpermission[1]{} 
\usepackage{xcolor}

\newcommand{\revision}[1]{\textcolor{black}{#1}}
\newcommand{\sid}[1]{{#1}}

\makeatletter
\def\@copyrightspace{\relax}
\makeatother

\begin{document}

\title{Designing Composites with Target Effective Young's Modulus using Reinforcement Learning}

\author{Aldair E. Gongora}
\authornote{Authors contributed equally to this research.}
\email{agongora@bu.edu}
\author{Siddharth Mysore}
\authornotemark[1]
\email{sidmys@bu.edu}
\affiliation{%
  \institution{Boston University}
  \streetaddress{Boston University}
  \city{Boston}
  \state{Massachusetts}
  \country{USA}
  \postcode{02215}
}

\author{Beichen Li}
\authornotemark[1]
\author{Wan Shou}
\author{Wojciech Matusik}
\affiliation{%
  \institution{Massachusetts Institute of Technology}
  \streetaddress{Massachusetts Institute of Technology}
  \city{Cambridge}
  \state{Massachusetts}
  \postcode{02139}
  \country{USA}}

\author{Elise F. Morgan}
\author{Keith A. Brown}
\affiliation{%
  \institution{Boston University}
  \streetaddress{Boston University}
  \city{Boston}
  \state{Massachusetts}
  \country{USA}
  \postcode{02215}
}

\author{Emily Whiting}
\affiliation{%
  \institution{Boston University}
  \streetaddress{Boston University}
  \city{Boston}
  \state{Massachusetts}
  \country{USA}
  \postcode{02215}
}
\email{whiting@bu.edu}

\renewcommand{\shortauthors}{Gongora, A.E., Mysore, S., and Li, B., et al.}

\begin{abstract}
    Advancements in additive manufacturing have enabled design and fabrication of materials and structures not previously realizable. In particular, the design space of composite materials and structures has vastly expanded, and the resulting size and complexity has challenged traditional design methodologies, such as brute force exploration and one factor at a time (OFAT) exploration, to find optimum or tailored designs. To address this challenge, supervised machine learning approaches have emerged to model the design space using curated training data; however, the selection of the training data is often determined by the user. In this work, we develop and utilize a Reinforcement learning (RL)-based framework for the design of composite structures which avoids the need for user-selected training data. For a 5 $\times$ 5 composite design space comprised of soft and compliant blocks of constituent material, we find that using this approach, the model can be trained using 2.78\% of the total design space consists of $2^{25}$ design possibilities. Additionally, the developed RL-based framework is capable of finding designs at a success rate exceeding 90\%. The success of this approach motivates future learning frameworks to utilize RL for the design of composites and other material systems. 
\end{abstract}




\maketitle
\pagestyle{plain}

\section{Introduction}
    Engineered composite materials and structures are able to achieve superior mechanical performance in comparison to their individual constituents~\cite{Lakes1993,Ghiasi2009,Ghiasi2010,Dimas2013,Wegst2015}. 
    The ability to tailor their design to meet performance requirements has resulted in their widespread application in the aerospace, automotive, and maritime industries~\cite{Guoxing2004,Fratzl2007,Gu2016}. 
    While the design process has previously relied on domain expertise, bio-mimicry, brute force exhaustive search, or iterative trial and error~\cite{Yeo2018}, recent advancements in additive manufacturing (AM) have tremendously enhanced the realizable design space and challenged the conventional approaches to exploring design spaces~\cite{Libonati2016,Gu2017_PrintingNature,Gu2017_Hierarchy}. 
    The design freedom afforded by AM has significantly expanded the design space by enabling the fabrication of composites with arbitrary geometry and material distributions spanning various length scales. 
    With this expansion comes challenges such as how to rapidly and efficiently exploring the vast and complex design space for optimal or targeted mechanical performance.
    \sid{While more traditional optimization techniques have been proposed, their robustness is often limited by the complexity of the design problems.}
    
    \begin{figure*}[t]
    	\centering
        \includegraphics[width=0.8\textwidth]{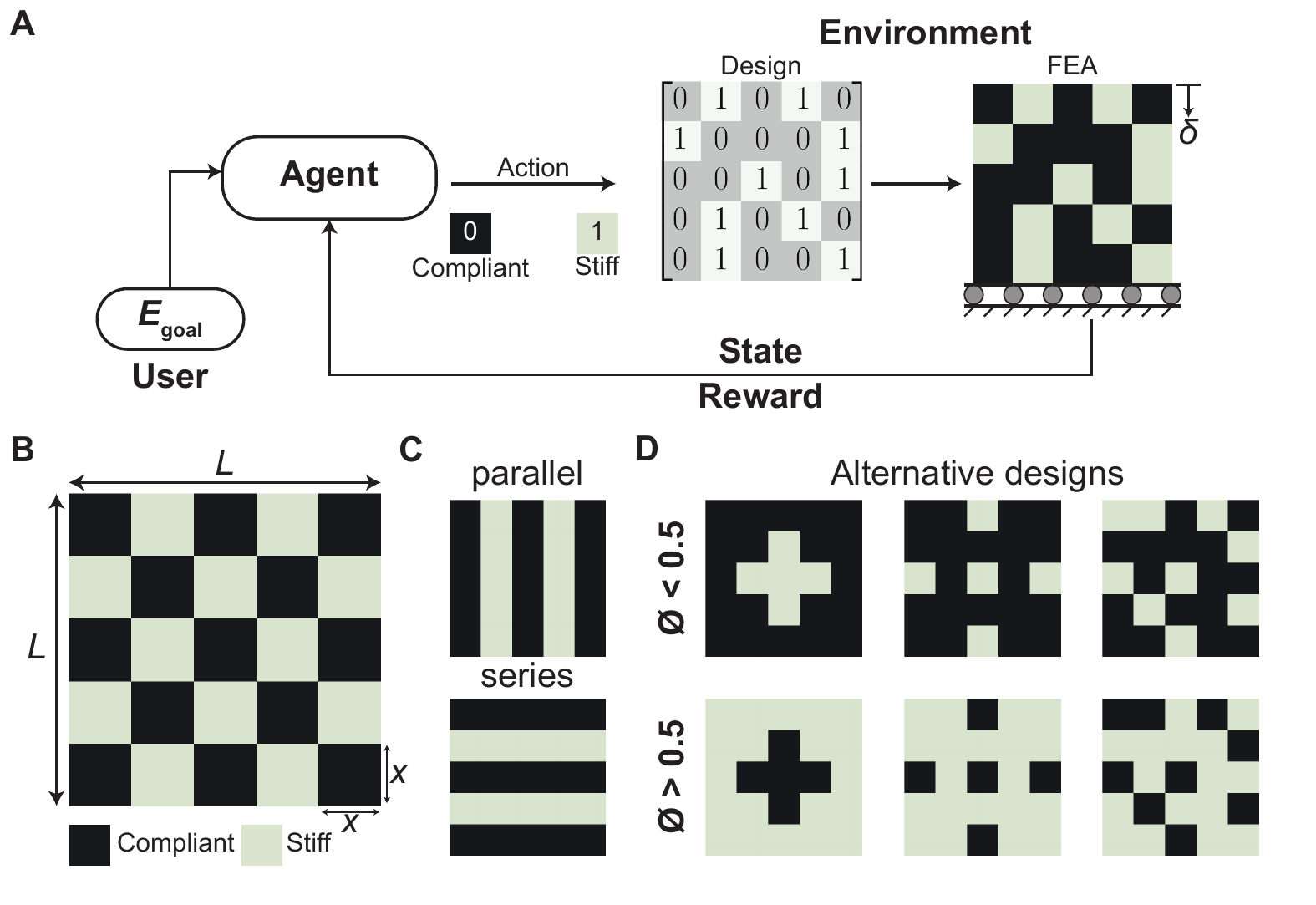}
     	\vspace{-1.5\baselineskip}
     	\caption{The Reinforcement Learning (RL) framework for the design of parametric composites consisting of stiff and compliant building blocks of constituent material to meet a specified target modulus $E_{goal}$ (A). The design space considered in this work is built from a $5$ by $5$ arrangement of blocks of side length $x$ where the composite sample has side lengths $L$ (B). The design space consists of parallel and series designs (C) with a volume fraction of stiff constituent material $\phi = 0.40$. The design space also consists of alternative designs where the majority of the compliant or stiff constituent material is concentrated at the center, sparsely distributed along the edges, or randomly distributed (D).}
        \label{fig:RLforCompDesignPipeline}
    \end{figure*}
    
    \sid{
    In order to overcome some of these design challenges, specifically those pertaining to exploring and modeling the design space, we propose using reinforcement learning (RL).
    RL algorithms learn to model the problem space through interaction and can be optimized to solve specific controls problems.
    }
    \sid{We} present a design framework that utilizes RL to design and find composite designs that meet a specified performance target~(Figure~\ref{fig:RLforCompDesignPipeline}~A).
    \sid{While the models learned do not provide a general solution for arbitrary composite design problems, our work offers a guideline on successfully framing design problems as RL problems.
    While learning-based frameworks can be more data and computationally intensive than any one optimization run using more traditional techniques, they offer potential benefits at scale, where inference after training can be significantly less expensive.}
    
    \sid{In this work, we mainly consider a 2-material composite design problem where we attempt to optimize the composite design in order to achieve specific material properties.}
    \sid{The} RL-based design framework, \sid{takes an initial composite design and user-specified desired material properties, comprising of target elastic properties and material composition, and iteratively modifies the design until the desired properties are satisfied.
    The design is parameterized as a 2D grid of constituent materials and the learned policy adjusts the design one grid cell at a time.
    This naturally upper-bounds the number of required interactions by the number of cells in the design grid.
    We demonstrate that, despite only exploring less than 5\% of the design space for our design problem during training, the learned RL policies are able to successfully solve over 95\% of the design tasks in testing.
    }

\section{Related Work}
    To overcome the challenges of design optimization in additive manufacture, techniques combining computational methods and optimization algorithms, such as topology optimization, have enabled the design of tailored composites with targeted structural and material design requirements such as elastic properties, Poisson's ratio, and tunable stress-strain curves~\cite{Sigmund1998,Sigmund2013,Gu2016_OptCompositeFractureProp,Zhu2017,Chen2018}.
    Although these approaches have achieved success in specific classes of problems, they are often limited by the complexity of the design space and associated computational cost.
    
    Recently, machine learning (ML) based design frameworks have achieved success in the design and discovery of materials and structures with optimal or targeted properties~\cite{Nikolaev2016,Gu2018_Composites,Chen2019,Abueidda2019,Mao2020,Gongora2020a,gongora2021PIBEAR}.
    Specifically in the field of composite design, ML techniques utilizing artificial neural networks and deep learning have been used in classification applications~\cite{Chen2020} and in inverse design applications~\cite{Yang2020}.
    These applications have demonstrated the ability to train ML-based models with a fraction of observations from the design space to successfully assess or predict mechanical performance. 
    A non-trivial challenge that faces the development of accurate ML models is the selection of appropriate training data, as inferior models can emerge from training on insufficient or poorly selected data. 
    In practice, the training data is often selected using uniform random sampling or a design of experiments approach, such as Latin hyper-cube sampling; however, the applicability of these approaches are limited by the size of the design space and the computational or experimental cost~\cite{Jin2020,Silver2017}.
    
    The dependence of the success of a ML model on appropriately selected training data motivates the development and application of an approach that automatically selects the training data. 
    An approach that can address this challenge is reinforcement learning (RL)~\cite{Popova2018,Andrychowicz2017,Botvinick2019,Leinen2020,Mnih2015}.
    RL models, also called RL agents, iteratively interact with a system in order to affect the state of the system. 
    In lieu of curating a dataset of behaviors, a reward function can be defined to reflect the quality of actions taken by agents. 
    The behavior of RL agents is conditioned to maximize the received reward. 
    This shifts some of the burden from dataset curation to reward function engineering, but the latter can be significantly more intuitive and tractable when dealing with large data-spaces.
    

\begin{figure*}[t]
	\centering
    \includegraphics[width=0.8\textwidth]{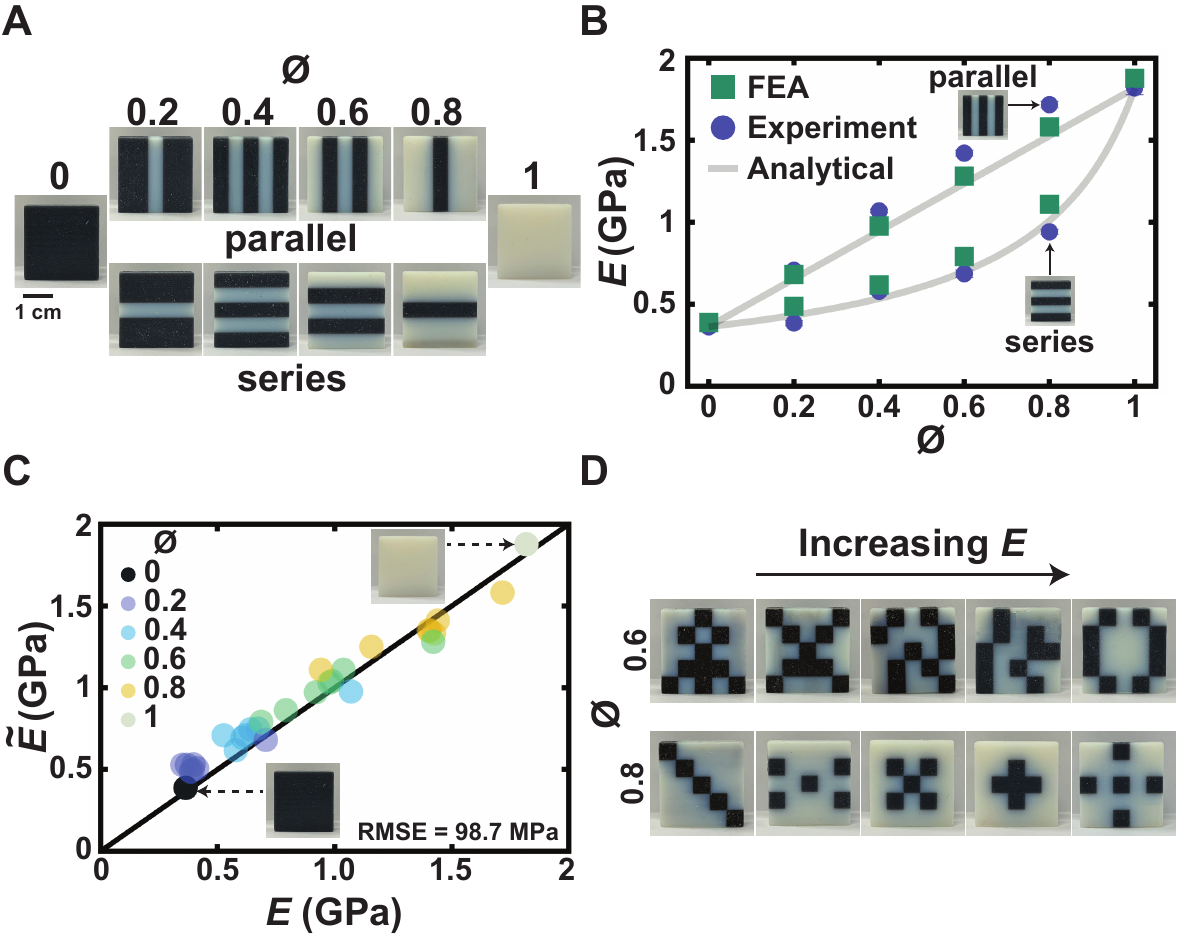}
 	\caption{Fabricated composite designs with volume fraction between $\phi = 0$ and $\phi = 1$ with classical parallel and series composite designs with $\phi = 0.2, 0.4, 0.6,$ and $0.8$ (A). Comparison of experimentally measured effective Young’s modulus $E$, FEA predicted effective Young’s modulus $\tilde{E}$, and analytical predictions of effective Young’s modulus for parallel and series composite designs with varying $\phi$ (B). Comparison of $E$ and $\tilde{E}$ for previously tested designs and five alternative designs for each $\phi = 0.2, 0.4, 0.6,$ and $0.8$ (C). Example designs with $\phi = 0.6$ and $\phi = 0.8$ in ascending order based on $E$ (D).}
    \label{fig:ExpSimValidation}
\end{figure*}

\section{Design Space and Finite-element analysis}
    In order to frame composite design as a reinforcement learning problem, we needed to establish a suitable state-space representation and build an environment for the RL algorithm with which to interact with.
    Training in simulation allows for faster feedback than fabrication and physical analysis, so we use finite element analysis (FEA) as our primary analytical tool for estimating design properties.
    Before employing the RL algorithm, it is imperative to validate the finite element predictions of Young's modulus $E$.
    This section discusses the composite design space considered and the development and validation of FEA pipeline.

\subsection{The Composite Design Space}

    Given that the set of all possible designs is large and intractable, we consider here a more constrained problem amenable to experimental validation so as to establish a viable framework for RL-automated design. 
    We consider a composite design space built from a $5 \times 5$ binary arrangement of material blocks where each block can be composed of one of two materials that differ in their elastic moduli. The stiffer of these two materials we term "stiff" and the more compliant of these materials we term "compliant".
    The material blocks of the composite have a side length $x = 5$ mm while the total composite has a side length $L = 25$ mm and depth of $5$ mm~(Figure~\ref{fig:RLforCompDesignPipeline}~B). 
    Without considering any geometric symmetries, a total of $2^{25}$ distinct designs exist in the composite design space. 
    Classical parallel and series composite designs can be found in the design space by varying the volume fraction $\phi$ of stiff material ~(Figure~\ref{fig:RLforCompDesignPipeline}~C). 
    Alternative designs also exist;
    for instance, designs where the majority of the compliant or stiff constituent material is concentrated at the center, sparsely distributed along the edges, or randomly distributed~(Figure~\ref{fig:RLforCompDesignPipeline}~D).
    With the ability to fabricate alternative designs, a range of Young's modulus values can be achieved for a given volume fraction.

\subsection{Development and Validation of a Predictive Model using Finite Element Analysis}
    
    The model we developed uses 2-D explicit FEA,which was implemented in C++ using the Taichi Graphics Library~\cite{hu2018taichi}, to predict the effective Young's modulus of the composite in the $5 \times 5$ design space. Here, we used a Neo-Hookean model for both the compliant and stiff constituent materials and a static compressive strain of 1e-4.
    The stiff and compliant constituent materials considered in this work were VeroWhitePlus (VW+) and a volume percentage mixture of 50\% VW+ and 50\% TangoBlackPlus (TB+), respectively. 
    \revision{The effective Young's modulus of the composite was estimated from the predicted stress at a prescribed compressive strain of 1e-4, which was kept constant for all simulations in this study. 
    The effective Young's modulus was calculated by dividing the predicted stress by the compressive strain. 
    To simulate the uniaxial compression of the composite design, displacement boundary conditions were imposed on the nodes of the top surface of the composite.}
    The coefficients of the Neo-Hookean model were computed from the Young's modulus and Poisson's ratio of the constituent materials. 
    The Young's modulus and Poisson's ratio of the stiff constituent material was $1818$ MPa and $0.33$, respectively, while the Young's modulus and Poisson's ratio of the compliant constituent material was $364$ MPa and $0.49$, respectively. 
    \revision{Additionally, an explicit solver was selected over an implicit solver in the FEA pipeline to avoid potential computational bottlenecks arising from the need to  invert the stiffness matrix of the course of the applied deformation in implicit analysis. 
    In the explicit solver, the time step was set to 2.3e-7, which was sufficiently small to prevent numerical instability in forward Euler integration. 
    A fixed 2000 iterations was used in the solver which was determined to be sufficiently large to guarantee convergence without compromising computational time based on preliminary tests. 
    Furthermore, the parameters used in the solver were determined to be appropriate based on the reasonable agreement observed between Young modulus predictions and experimental measurements.}

    The selected composite designs in this study were fabricated from the aforementioned stiff and compliant constituent materials and experimental measurements were taken in triplicate for each composite design where $E$ is the mean effective Young's modulus of the measurements.
    To assess the model using FEA, the predicted effective Young's modulus $\tilde{E}$ of selected composite designs were compared to experimental measurements of the effective Young's modulus $E$ from quasi-static uniaxial compression tests conducted on a universal testing machine (Instron 5984). 
    The first set of selected composites designs were designs with only compliant constituent material $\phi = 0$, only stiff constituent material $\phi = 1$, and classical parallel and series composite designs with $\phi = 0.2$, $0.4$, $0.6$, and $0.8$ (Figure~\ref{fig:ExpSimValidation}~A). 
    The absolute relative error between $\tilde{E}$ and $E$ for composite designs with $\phi = 0$ and $\phi = 1$ were $6.97$\% and $3.35$\%, respectively, while the average absolute relative error between $\tilde{E}$ and $E$ for the classical parallel and series composite designs with $\phi = 0.2$, $0.4$, $0.6$, and $0.8$ was $12.10$\% (Figure~\ref{fig:ExpSimValidation}~B). 
    Evaluating the entire dataset with $\phi = 0$, $0.2$, $0.4$, $0.6$, $0.8$, and $1.0$, the root mean squared error (RMSE) was 101.79 MPa with an $R^2=0.96$ where $R$ is the Pearson correlation coefficient. 
    For these composite designs, $E$ can also be analytically estimated via the Voigt and Reuss approximations for parallel and series designs, respectively. 
    The approximations also yielded reasonable predictions of $E$ with an average absolute relative error of $9.07$\% when compared to experimental measurements. 
    While the Voigt and Reuss approximations enable rapid predictive approximations of $E$, the models are only expected to be accurate for parallel and series composite designs.
    
    To further assess the predictions using FEA, the $\tilde{E}$ of five alternative designs for each $\phi = 0.2$, $0.4$, $0.6$, and $0.8$ were compared to experimental measurements from uniaxial compression testing. 
    Adding these results to the classical parallel and series dataset, the agreement between $\tilde{E}$ and $E$ was evaluated.
    The RMSE was computed to be $98.69$ MPa with $R^2=0.97$ and the average absolute relative error between $\tilde{E}$ and $E$ of all the selected designs was calculated to be 13.8\% (Figure~\ref{fig:ExpSimValidation}~C).
    Based on these results, we concluded that the predictions of Young's modulus from FEA were in good agreement with experimental measurements. 
    Interestingly, from the E of alternative designs with varying volume fraction, a broad range of achievable modulus values with non-trivial designs were observed. 
    Specifically, for $\phi=0.6$ the minimum $E$ was 792.3 MPa and the maximum $E$ was approximately 31\% larger at 1.04 GPa.
    Additionally, for $\phi=0.8$, the minimum $E$ was $1.16$ GPa and the maximum $E$ was approximately $24$\% larger at $1.44$ GPa (Figure~\ref{fig:ExpSimValidation}~D).
    While this range was observed for the experimental measurements of alternative designs, brute-force evaluations of all the designs suggest that for a given $\phi$ there is a range of achievable $E$ via alternative designs.
    The overlap in ranges for varying $\phi$ demonstrates the ability to vary performance for a given $\phi$, which is a key factor in designing composites to meet performance requirements. 
    \revision{Note that $\tilde{E}$ for a composite design only corresponds to quasi-static uniaxial compression conditions. 
    While designs such as the alternative composite designs may possess $E$ that depend on the loading direction, only quasi-static compression conditions as shown in Figure~\ref{fig:RLforCompDesignPipeline}~A are considered in this study.} 
    
    \begin{figure*}[h]
    	\centering
        \includegraphics[width=0.8\textwidth]{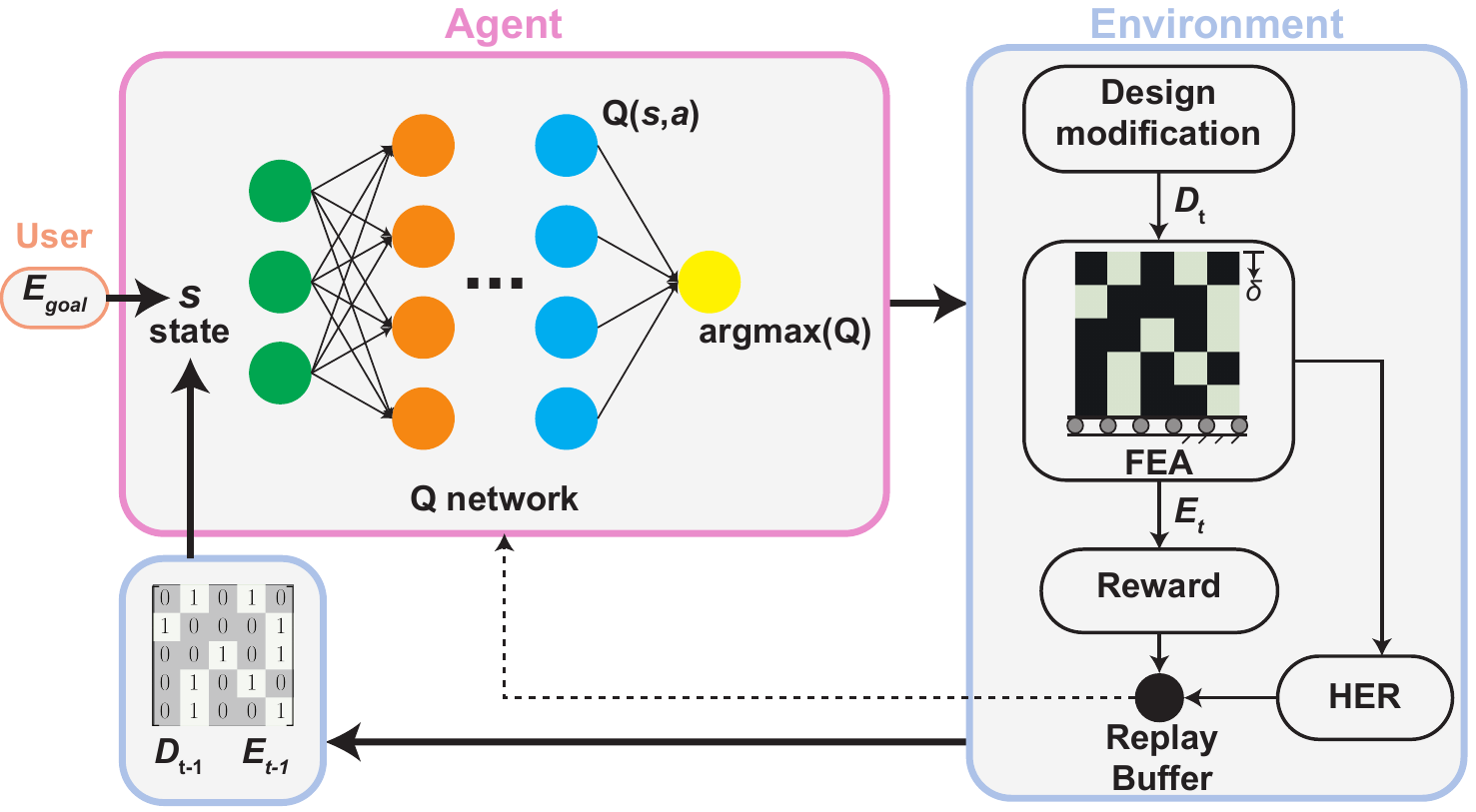}
     	\caption{Overview of the reinforcement learning (RL) for composite design framework}
        \label{fig:RLframework}
    \end{figure*}

\section{Developing the Reinforcement Learning (RL) Framework for Composite Design}
    Reinforcement learning (RL) algorithms seek to solve a sequential decision-making problem, which are typically formulated as a Markov decision process (MDP). 
    There are two major components of any RL setup – (i) the training environments, and (ii) the agent(s), i.e., the learned behavior model. 
    Agents take actions in response to the current state, $s_t$, of the environment and the state of the environment changes in response to the action taken. 
    The agent decides its actions by attempting to maximize the cumulative reward, received from the environment. 
    In order to successfully train an RL agent on a task, it is important to balance the complexities of the agent policy as well as the task environment. 
    While more complex policy functions can have an increased modeling capacity which may allow learning of more complex tasks, this comes with increased computational costs and training data requirements. 
    We designed our task environment with this in mind.
    
    Figure~\ref{fig:RLframework} provides an overview of the RL framework developed to design composites with a target Young's modulus $E_{goal}$ and volume fraction $\phi_{goal}$. 
    We began by considering similar problems already solvable by RL and found close parallels with the problem of bit-flipping – i.e., modifying one binary string to match another given one, one element at a time. 
    Like in bit-flipping, our design problem includes a current state and a desired future state, where our 2-material design, $D_t$, can be represented as a binary string (where an $n \times m$ grid can be represented by a nm-dimensional binary vector). 
    Unlike the bit-flipping problem; however, we do not know the desired design ahead of time, only the desired properties. By including the current (at iteration $t$) and desired material properties, $E_t$ and $E_{goal}$, respectively, in the state representation we equivalently indicate necessary information about a goal state to allow the agent to model and achieve desired material properties. 
    Formulating the design problem as a sequential task in this way enables two key advantages: (i) it maintains the Markovian properties of states being independent of their histories, thus allowing RL to solve this problem as an MDP, and (ii) it allows for a natural representation of termination state and computational limit and consequently lends itself to an intuitive reward signal.
    
    \begin{figure*}[h]
    	\centering
        \includegraphics[width=0.8\textwidth]{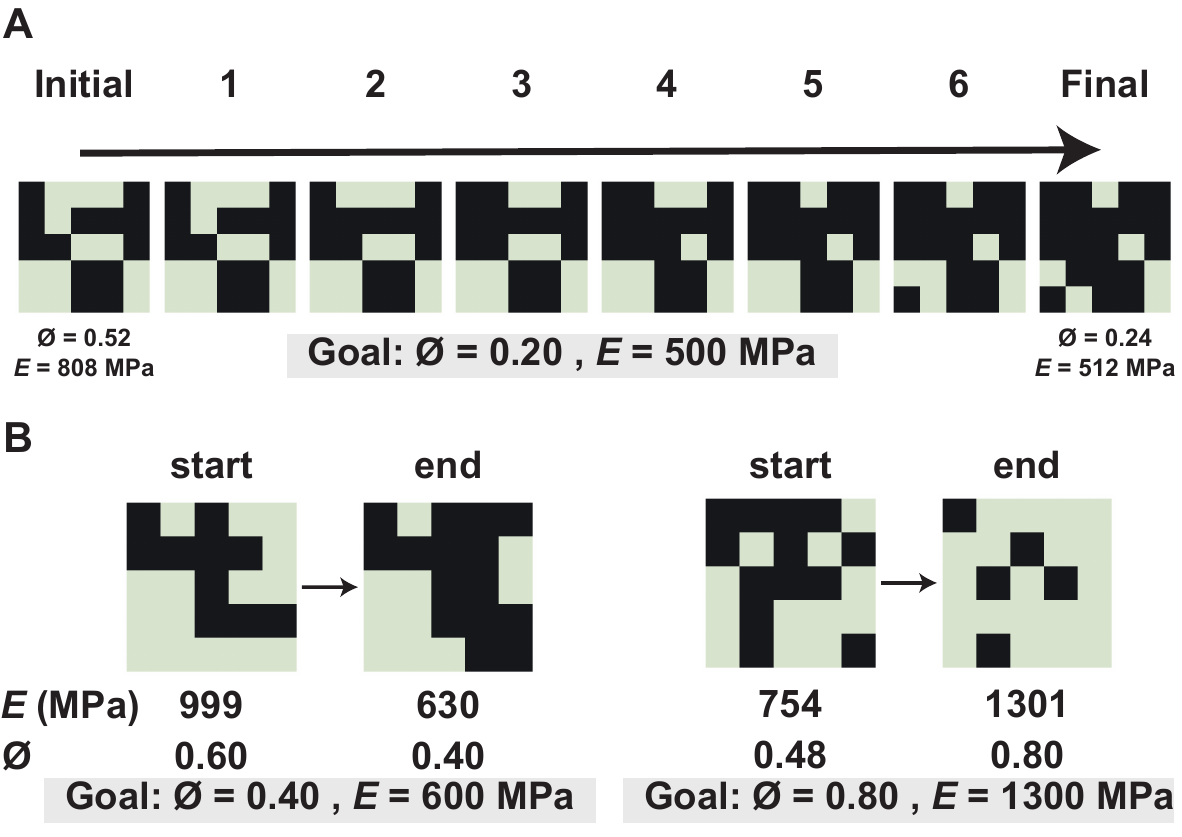}
     	\caption{Sequential design changes from a randomly selected initial to final design made by a trained agent to meet a goal $E = 500$ MPa and goal $\phi = 0.2$ (A). Two examples with randomly selected initializations and the final designs returned by the trained agent (B).}
        \label{fig:SeqDesignExample}
    \end{figure*} 
    
    Defining the design problem as a sequential RL problem allows us to side-step difficulties that would normally arise with procuring a training set for supervised learning.
    In a typical supervised learning approach for a design problem such as this, or indeed any search problem, one of the key issues would be in determining the `quality' of arbitrary states with respect to a desired goal state.
    Computing the scalar difference between the desired and current material properties does not directly translate to a strategy on traversing the design space towards a viable design that satisfies the design goals, prohibiting direct optimization.
    Without a clear notion of distance, or first knowing what an ideal search trajectory is, it would be difficult to quantify the quality of any given state, thus making it difficult to define loss signals to facilitate training.
    In contrast, RL allows the use of a significantly more intuitive binary reward signal to guide training: 
    if the properties of the current design are within a tolerance of the desired properties, the agent gets a zero-reward, and it gets a negative reward $(-1)$ in all other cases.
    It is then left to the RL algorithm to learn to evaluate how to best traverse the manifold of the design space in order to satisfy design requirements.
    
    Our agents are represented by Q-networks, which gained prominence when \cite{Mnih2015} demonstrated the viability of deep RL in learning to play a diverse set of video-games. 
    Video-games require agents to achieve and maintain desirable states, and this is typically achieved through sequential interaction, much like the design problem we consider here. 
    The Q-network learning framework is developed as an extension of the tabular Q-learning problem \cite{Watkins1992,RL_Book2018}, where the RL model learns a Q-value function mapping states and actions to expected rewards. 
    If an agent is able to accurately model the Q-value, it can optimize its actions at every state to maximize rewards. 
    Deep Q-networks extend tabular Q-learning by representing the Q-function not as a table enumerating all states and actions but rather a neural network that predicts the Q-value for any possible action given an input state. 
    The agent is also given a strict limit of $nm$ iterations to solve the problem for an $n \times m$ grid – as the maximum design distance between a desired design and the current one (or any arbitrary two designs) cannot exceed the number of material cells. 
    It naturally follows \revision{from the Q-lerning optimization} that the best way for the agents to learn to maximize rewards is to solve the problem as quickly as possible.
    
    While suitable training environment and RL agent design enable effective learning, they still may not be efficient enough to be practical as deep RL can be highly data intensive \cite{Leinen2020}.
    Given the vast number of variations in the configuration space and the not insignificant time that it takes to analyze a composite with FEA, it was important to improve data efficiency during training. 
    To that end, we utilize Hindsight Experience Replay (HER) \cite{Andrychowicz2017}, a technique developed for goal based deep RL, where training data is synthetically augmented based on data gathered while agents explore their training environments. 
    Briefly, HER treats the final and/or intermediate states reached in a sequence of environment interactions as synthetic target goals.
    While the true goal may not be reached, the sequence of interactions still provides agents information on how to reach the states that were visited, should they become relevant in future.
    In effect, the virtual re-contextualization of prior experience allows every interaction with the environment to be several times more useful as a data-point, thus improving data efficiency.
    
    For our $5 \times 5$ grid designs, agents are trained in cycles of experience gathering and optimization over 750 cycles with 50 episodes per cycle, for a total of 37500 episodes. 
    Episodes are capped by the total number of cells in the material design (in our case 25) as this is the maximum number of changes that should be required for any design. 
    In all, agents make fewer than 1 million calls to the finite element analysis (FEA) solver, which, while a significant number, corresponds to sampling at most 2.8\% of the $2^{25}$ design options during training. 
    While the true unique design space is likely smaller when accounting for design symmetries, our method still offers relatively low data complexity for learning. 
    Further details on setup and training are provided in Appendix~\ref{Appendix:RL}.
    
    \begin{figure*}[t]
    	\centering
        \includegraphics[width=0.8\textwidth]{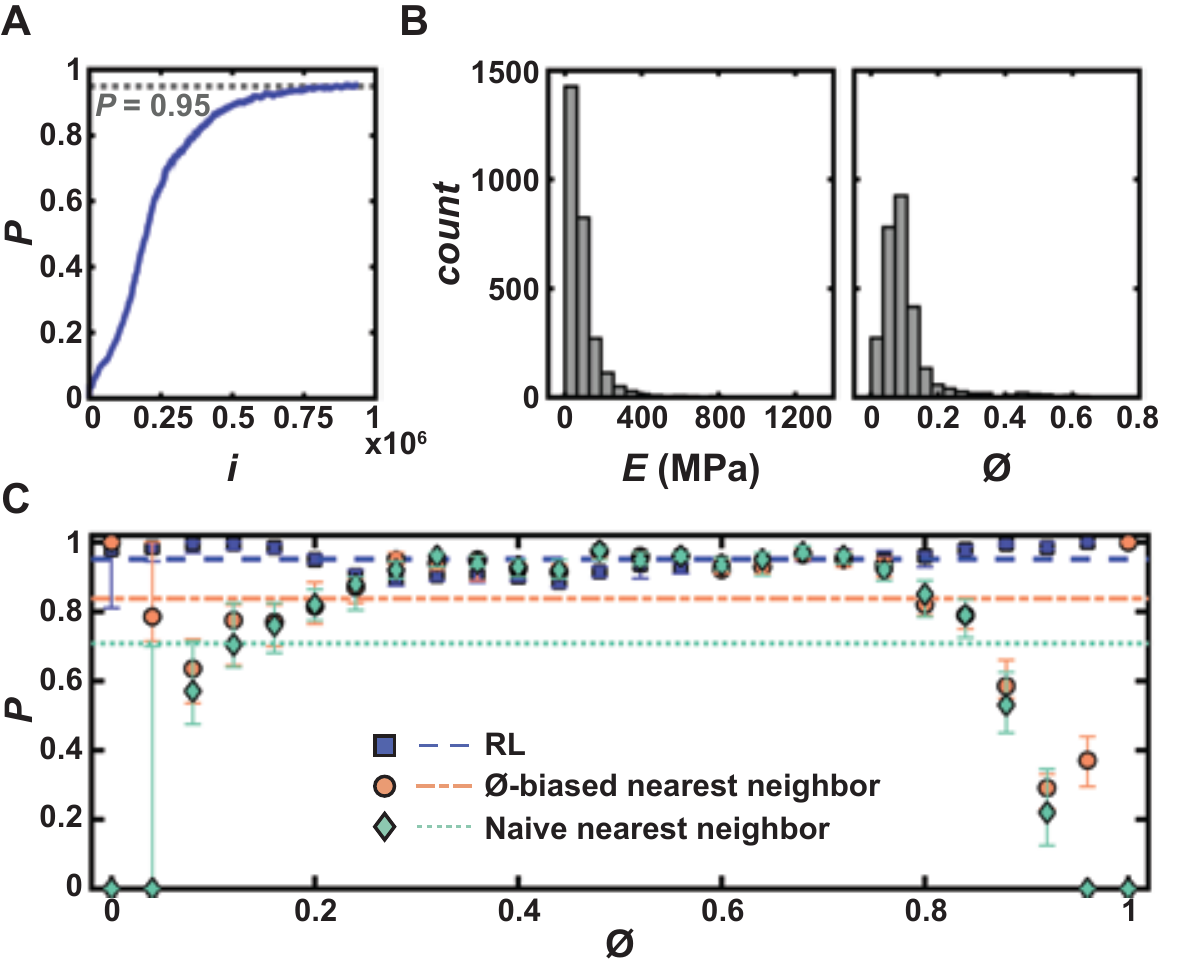}
        \vspace{-\baselineskip}
     	\caption{Median success rate $P$ as function of the number of environment interactions $i$ during training evaluated over $20$ independent agents with the shaded region represents the $25$th and $75$th percentiles (A). Distribution of deviation from desired properties in failure cases during testing (B). $P$ as a function of $\phi$ for the $52,000$ random tests (C).}
        \label{fig:SuccessStats}
    \end{figure*}


\section{Results and Discussion}

    To evaluate the performance of RL framework for composite design, we consider the previously described composite design space built from a $5 \times 5$ binary arrangement of materials blocks where each block can be either compliant or stiff material. 
    The user specifies a target Young's modulus $E_{goal}$ and volume fraction $\phi_{goal}$ and the trained agent has a maximum budget of 25 iterations to return a final design -- corresponding to the total number of cells in the grid.
    
    For example, given $E_{goal} = 500$~MPa, $\phi_{goal} = 0.20$, and a randomly selected initial design configuration with an initial $E_0 = 808$ MPa and $\phi_0 = 0.52$, the agent returns a design with final $E = 512$~MPa and $\phi = 0.24$ in seven sequential design changes (Figure~\ref{fig:SeqDesignExample}~A).
    The absolute relative error in $E$ is 2.4\%.
    Due to the discrete nature of the $5 \times 5$ design space, the error in $\phi$ was the addition of one stiff material block which resulted in the final $\phi$ being larger than the target $\phi$ by 0.04.
    Due to the number of possible design variations for volume fractions $\phi \in [0.24,0.48]$, we observed that agents were typically least performant in this region of the design space.
    Additionally, two examples are shown in Figure~\ref{fig:SeqDesignExample}~B where one example, Figure~\ref{fig:SeqDesignExample}~B1, has a $\phi_{goal} = 0.4$ and $E_{goal} = 600$~MPa and another example, Figure~\ref{fig:SeqDesignExample}~B2, where $\phi_{goal} = 0.80$ and $E_{goal} = 1300$~MPa. 
    For the example in Figure~\ref{fig:SeqDesignExample}~B1, the final returned configuration exactly satisfies $\phi_{goal} = 0.4$ and slightly exceeds $E_{goal} = 600$~MPa by 5\%. 
    In the second case, Figure~\ref{fig:SeqDesignExample}~B2, the returned configuration exactly satisfies $\phi_{goal} = 0.8$ and slightly exceeds $E_{goal} = 1300$~MPa by 0.08\%.
    The agent begins with a randomly selected initial design and it then finds designs that meet both specified goals with an absolute relative error less than or equal to 5\%. 
    From these examples, it is observed that agents have preserved aspects of the original design while making modifications to achieve the user specified goal. 
    The observed behavior is indicative that the agents have not just learned to memorize solutions but to modify designs to achieve desired properties while under the prescribed budget, and ideally with the fewest changes.
    This follows as a natural result of the RL optimization criteria however, as the optimal strategy that maximizes learning rewards is the one that is able to satisfy the design problem in the fewest iterations.
    
    While the previous results are promising, they represent a single trained agent.
    To further assess the performance of the RL framework for the design of composites, we tested our learning framework with 20 independently trained agents over which we achieved a consistent and median success rate $P$ of 95\% (Figure~\ref{fig:SuccessStats}~A). 
    The 95\% median success rate reflects the average fraction of 2600 of design problems successfully solved by each agent (a total of 52,000 randomly conducted tests) within desired tolerances. 
    This is achieved while requiring exploration of only 3\% of the \revision{design} space – aided significantly by using HER.
    Notably, there is a low variance in the performance among agents reflecting the stability of the RL framework in designing composite.

    While approximately 5\% of cases are categorized as ``failed", they did not fail catastrophically but rather failed to achieve a design within the desired tolerances within 25 iterations (Figure~\ref{fig:SuccessStats}~B). 
    Specifically, the average error of $E$ of the agents was $61 \pm 8$ MPa while the average error of $\phi$ was $8 \pm 2$\%. 
    In all cases, the tolerances for success were set at $E = 50$ MPa and $\phi = 0.04$ (corresponding to a 4\% error in the volume fraction). 
    The tolerance enabled the agents to rapidly locate designs that met specified user requirements in the vast design space. Additionally, considering the prescribed tolerances, even in cases of `failure' the error in the final model's properties marginally outside the tolerance thresholds. 
    While average $P$, is an important metric, the variation of $P$ as a function of $\phi$ can also yield insight into the performance of the RL framework (Figure~\ref{fig:SuccessStats}~C).
    \revision{
    We also compare the RL approach to nearest neighbor based solutions based on sampling the design space using as many samples as the upper bound of unique configurations visited by the RL approach (i.e. 2.78\%) (details in Section~\ref{NNDisc}).
    }
    For $\phi$ in the range of 0.24 and 0.6, $P$ is observed to slightly fall below 0.90. Specifically, $\phi = 0.48$ has the lowest median $P = 0.89$. The decrease in $P$ reflects that this area in the design space possesses the largest number of possible configurations. 
    \revision{
    When comparing this against the nearest neighbor approaches, we see that the RL agents do perform marginally worse, on average, in the regions of the design space where there are many possible design configurations.
    This was to be expected, since sampling from the design space, either uniformly at random, or randomly within the possible designs for each possible $\phi$, would result in a large proportion of the total samples being in the range of $\phi \in [0.4, 0.6]$, as this $\phi$-range represents over 77\% of the design space.
    Crucially though, we see that the RL method is able to maintain relatively consistent levels of $P$ even in the regions of design space that are represented by fewer possible design configurations. 
    Since the RL-based solver effectively builds its own internalized model of the system, it can compensate for the relative lack of representation in the training data.
    In contrast, highly data-dependent techniques, such as nearest neighbor search, would fail to generalize outside of the available data, allowing the RL approach to to achieve a higher average $P=0.95$ as compared to $P=0.84$ and $P=0.71$ for the $\phi$-biased and naive nearest neighbor searches, respectively.
    Furthermore, the RL approach tries to maintain design similarity to the original input, which would be difficult to achieve through sampling-based approaches, given their data reliance.
    }
    

\section{Conclusion}
    In this work, we developed and utilized a RL-based framework for the design of composite structures to meet user-specified target Young's modulus values E. 
    The effectiveness of this approach was evaluated using a $5 \times 5$ composite design space that was comprised of stiff and compliant constituent materials. 
    Using the RL-based approach, the models were trained using approximately 2.78\% of the $2^{25}$ design space. 
    Tested on a total of 52000 test cases, 20 independently trained RL agents successfully solved the design optimization problem in 95\% of the tests. 
    In totality, this work demonstrated the promise of RL in materials design since traditional design of experiment approaches are limited by the size of the design space and supervised machine learning approaches are limited by the quality of the training data. 
    While we recognize that there are still open questions on the extensibility of such an approach to more complex designs and design requirements, this work is meant to establish a framework and the viability of RL applied to automated design. 
    \revision{Additionally, the results described in this work demonstrate the viability of RL for composite design motivating future work and applications considering the distribution and arrangement of multiple configurations of composite designs composed of material blocks. Notably, in these cases, size and boundary effects become imperative to consider in tandem with designing the RL framework for a larger design space.}
    An RL-based design framework circumvents the challenges faced by traditional design of experiment approach and supervised ML approaches by being able to automatically generate its own training set and solve subsequent design problems. 
    This further motivates design approaches in materials design and discovery to focus on exploring larger design spaces.

\begin{acks}
    This material is based upon work supported by the National Science Foundation under Grant No. 1813319, the Alfred P. Sloan Foundation: Sloan Research Fellowship, Google LLC, the Boston University Rafik B. Hariri Institute for Computing and Computational Science and Engineering (2017-10-005), and the U.S. Army CCDC Soldier Center (contract W911QY2020002). 
\end{acks}

\vspace{\baselineskip}

\bibliographystyle{ACM-Reference-Format}
\bibliography{references}

\newpage
\appendix

\section{Materials Modeling, Fabrication and Analysis Methods}\label{Appendix:Fab}

\subsection{Fabrication and Mechanical Testing}

    All composite specimens used in this study were printed using a multi-material 3D printer (Objet260 Connex). 
    The specimens were printed using two constituent materials. The first constituent material was VeroWhitePlus (VW+) and the second constituent material was a volume percentage mixture of 50\% VeroWhitePlus and 50\% TangoBlackPlus (TB+). 
    Compression tests were conducted using a universal testing system (Instron 5984) at a loading rate of 3 mm/min. 
    Three samples for each selected composite design were fabricated and tested to obtain a mean measurement of Young's modulus, which was measured as the slope of the linear portion of the tested specimen's stress-strain curve. 

\subsection{Analytical analysis}

    The achievable range of Young's moduli $E$ of a composite design comprised of two constituent materials is bounded by the Voigt and Reuss analytical models. 
    The Voigt model assumes that composite designs in parallel can be modelled as springs acting in parallel and the Reuss model assumes that the composite designs in series can be modelled as springs in series. 
    The models are a function of the stiff constituent's volume fraction $\phi_s$, the modulus of the stiff constituent material $E_s$ and the modulus of the compliant constituent material $E_c$. 
    The Voigt model can be expressed as $E = \phi_s E_s + (1 - \phi_s ) E_c$. 
    The Reuss model can be expressed as $E = (E_c E_s) / (\phi_s E_c + (1 - \phi_s ) E_s)$. 
    The modulus of the constituent materials $E_s$ and $E_c$ are 1,818 MPa and 364 MPa, respectively.

\subsection{Finite-element analysis}

    \revision{In order to predict the $E$ of the composite designs, we performed 2-D finite-element analysis (FEA). 
    The $5 \times 5$ composite design was represented using a $40 \times 40$ grid of quad-finite elements where each cell of the composite was represented by an $8 \times 8$ grid. 
    We adopted a Neo-Hookean material model to compute the hyper-elastic response of each element, where the strain energy density function is given by~\cite{sifakis2012fem}.
    \begin{equation}
        W = \frac{\mu}{2}\left(I_1 - 2 - \ln J\right) + \frac{\lambda}{2}\left(\ln J\right)^2.
    \end{equation}
    Here, $\mu$ and $\lambda$ are the Lam\'{e} parameters; $I_1$ is the first invariant of the right Cauchy-Green deformation tensor; $J$ is the determinant of the deformation gradient.
    The material coefficients of the stiff and compliant base materials ($\mu_s, \mu_c, \lambda_s, \lambda_c$) were computed from Young's moduli ($E_s,E_c$) and Poisson's ratios ($v_s,v_c$) as follows~\cite{kelly2013solid}.
    \begin{align}
    \begin{split}
        \mu &= \frac{E'}{2(1 + \nu')}, \\[0.25em]
        \lambda &= \frac{E'\nu'}{(1 + \nu')(1 - 2\nu')}, \\[0.25em]
        \nu' &= \frac{\nu}{1 + \nu}, \\[0.25em]
        E' &= \frac{E(1 + 2\nu)}{(1 + \nu)^2}.
    \end{split}
    \end{align}
    We set $E_s$, $E_c$ the same as experimental measurements with $v_s = 0.33$ and $v_c = 0.49$.
    Additionally, we applied negative exponential damping to the nodal velocities ($v_{i,j}$) to address material viscosity in the FEM. 
    \begin{equation}
        v'_{i,j} = v_{i,j}e^{-\gamma \Delta t_\text{sim}}.
    \end{equation}
    The damping constant ($\gamma$) was set to 1e5 which leads to desirable damping effects in constituent material simulations. 
    Based on these settings, the deformation was solved using an explicit solver under the Dirichlet boundary constraint of a static compressive strain equal to 1e-4 on the displacement in the loading direction. 
    The time step ($\Delta t_\text{sim}$) was set to 2.3e-7 which is sufficiently small to prevent numerical instability in forward Euler integration. 
    Above all, we ran a fixed 2,000 iterations to guarantee convergence. The estimated Young's modulus was then derived from the gauge stress measured at boundary nodes. 
    The simulator was implemented in C++ with Taichi Graphics Library~\cite{hu2018taichi}. Furthermore, we note that the simulator is not the bottleneck of our computational pipeline, since each simulation takes well less than a second on a single CPU core.}
    
\section{Reinforcement Learning Framework}\label{Appendix:RL}

\subsection{Task Environment Design}

    At \revision{design-step} $t$, the task environment state is represented numerically by a 29-dimensional $s_t \in S$ where $s_t^i \in [0,1]$ and $S \subset R^{29}$. 
    The first 25 components capture the $5 \times 5$ material design grid, $D_t$. 
    The 26th and 27th components capture the current Young's modulus and volume fraction of material composition, while the 28th and 29th components capture the desired modulus and volume fractions respectively. 
    Neural networks are known to be sensitive to the scale of the inputs, so we scale the Young's moduli by dividing it by the maximum value of the two materials' moduli, i.e. $f(E) = \frac{E}{\max{(E_1 ,E_2)}}$.
    The action taken by the agent, $a_t$, in effect, selects a material cell in the design grid to `flip' between the two materials. 
    The resultant design matrix is evaluated for its material properties and this, along with the modified design is returned as the next state, $s_{t+1}$, to the agent. 
    During training, $s_{t+1}$ is also used to compute the binary reward of $-1$ or $0$ depending on whether the state is within tolerance of the desired properties. 
    We use a Young's modulus tolerance of 50 MPa and a volume fraction tolerance of 4\%, i.e. being one material cell off from the desired composition.
    
    For each episode of the training run, which is limited to 25 iterations as that is the maximum number of changes that could theoretically be required to achieve a goal for a $5 \times 5$ grid, a goal volume fraction and desired Young's modulus is selected randomly from the range of possible volume fractions and Young's moduli as defined by the Voigt and Reuss approximations (as shown in Figure~\ref{fig:ExpSimValidation}). 
    \revision{A design, chosen uniformly at random, is also instantiated and evaluated}. If the design's properties start too close to the goal properties, the goals are reset until the desired Young's modulus is at least 100 MPa away from the current design's modulus.
    
\subsection{Network Design and Training}
    
    In order to approximate the Q-function, we use a neural network with an input layer, 3 fully connected hidden layers and an output layer. The input layer has 29 input nodes to read in the 29-dimensional state provided by the training environment. 
    The first and second hidden layers have 128 and 64 neurons respectively with Rectified Linear Unit (ReLU) activations~\cite{Fukushima1980, NairHinton2010}. 
    These are passed to a 26-dimensional layer which estimates the Q-value of 26 possible actions (one for each material cell `flip' and including a null action where the agent does nothing). 
    The output action is determined by an argmax over the predicted Q-values to select action with the highest expected value. 
    We use a Q-value discount factor of $\gamma=0.99$ and the networks are trained with an Adam optimizer~\cite{Kingma2015} with a learning rate of $10^{-3}$, a batch size of 320 and a training buffer size of $2 \times 10^6$.
    Training is run over 750 cycles with 50 episodes per cycle, with each episode consisting of 25 environment interactions (episodes are not terminated immediately upon success so that agents learn to not deviate from a good design). 
    500 optimization steps are performed at the end of every cycle.
    
    To mitigate over-fitting and over-estimation in the Q network, we employ Double Deep Q-learning. 
    Double Q learning~\cite{HasseltHado2010_DoubleQLearning,vanhasselt2015deep} uses two networks in training instead of just one. 
    The networks start with equal weights but only one is updated consistently during training (the `main' or `live' network) while the other (`target') network is held fixed for set intervals.
    When estimating Q-values for the training loss, the target network is used to estimate the temporal difference error~\cite{RL_Book2018} and is updated with weights from the live network only periodically. 
    During training, we update the target network once per cycle with a 95\% interpolation factor on the weights between the target and live networks.
    
\subsection{Developing an Exhaustive Dataset}
    
    In order to enable quick testing and iteration on experimental design, we developed an exhaustive dataset of all with the $2^{25}$ possible designs and their associated FEA values. While any individual agent requires under 1 million data samples (less than 3\% of the design space), each agent may experience a different set of data, as is the nature of RL training. 
    Each FEA call takes time however and, in an academic setting, this was found to limit early development of our RL framework. 
    To mitigate this, we parallelized the FEA computation of every possible design to build a library of material properties for all the designs which could be looked up by the RL agents during training. 
    This ensured that we were not re-computing the FEA for any previously computed model and it also allowed us to massively parallelize the training of multiple agents when verifying the efficacy of our method, to demonstrate that the performance was consistent and repeatable. 
    We stress again that this was only done to facilitate the academic pursuit of a performance analysis of the method – any individual agent would only need to be connected directly to the FEA solver during training and would only need to be trained once.

\subsection{\revision{Testing against brute-force methods}}\label{NNDisc}

    \revision{
    The development of an exhaustive dataset also allowed us to compare our RL method against brute-force methods to solve a similar design problem.
    A naive brute-force algorithm which keeps with the bit-flipping inspired problem formulation would present with too much uncertainty in the run-time and could require a lot of sampling from the dataset - which would, in a real application, correspond to significant time spent on the FEM solver.
    Instead, we compare our method against a nearest-neighbor approach, where a subset of designs are sampled from the full set of possible designs and given design requirements, the nearest neighbor from the sampled subset is used to present a possible solution.
    The nearest-neighbor solver is easily bounded to use a similar amount of total data as the RL algorithm by sampling 3\% of the dataset.
    }
    
    \revision{
    We employed two sampling strategies for building the nearest-neighbor search subset: 
    (i) Naive Nearest Neighbor: using 3\% of the data, sampled uniformly from the full dataset, 
    and (ii) $\phi$-biased Nearest Neighbor: using 3\%, of the possible samples for each of the 26 possible volume fractions, totaling to 3\% of the full design space; both cases are rounded up to their nearest integer.
    The former is an admittedly naive approach but is likely more reflective of the sampling the RL agents might be expected to encounter, since we make no explicit attempt to condition how the RL algorithms explore the design space.
    The latter however is a minor change which uses a little bit of our prior knowledge of the design space to better shape the nearest-neighbor solution.
    Since we do not track the number of unique states visited by the RL agents during training, a comparison of sampling strategies on strategy alone is more difficult to judge, but we are able to compare the performances on the design tasks.
    Given that the nearest-neighbor approaches do not attempt to model the design space, we do not compare the similarity of the final solutions to any initial state, as that is not something this technique can be reasonably expected to account for in this particular design problem.
    }


\end{document}